\documentclass[aps,pre,floatfix,
 reprint,floatfix,
 superscriptaddress,
 amsmath,amssymb,
 aps,
 longbibliography,
]{revtex4-2}
\usepackage{graphicx}% Include figure files
\usepackage{dcolumn}% Align table columns on decimal point
\usepackage{bm}% bold math

\usepackage[mathlines]{lineno}% Enable numbering of text and display math
\usepackage{svg}
\usepackage[normalem]{ulem}  
\usepackage{xcolor}

\newcommand{\tw}[1]{\textcolor{black}{#1}}

\usepackage[pdftex,bookmarks,colorlinks,breaklinks]{hyperref} % PDF hyperlinks, with coloured links
\hypersetup{linkcolor=red,citecolor=blue,filecolor=dullmagenta,urlcolor=blue} % colored links
\hypersetup{pdftoolbar=true,pdfpagemode=UseNone,pdfstartview=FitH, colorlinks=true,extension=}

\usepackage{placeins} % Allows use of \FloatBarrier to control float placement
\usepackage{hyperref}% add hypertext capabilities

\begin{document}

% \preprint{APS/123-QED}

\title{Dynamic Synchronization of Driven Self-Oscillators -- Modeling and Experiment}

\author{Zhenwei Xu}
\email{zhenxu@ethz.ch}
\affiliation{CAPS Laboratory, Department of Mechanical and Process Engineering, ETH Zürich, 8092 Zürich, Switzerland}
\author{Ulrich Kuhl}%
\affiliation{CAPS Laboratory, Department of Mechanical and Process Engineering, ETH Zürich, 8092 Zürich, Switzerland}
\affiliation{Université Côte d’Azur, CNRS, Institut de Physique de Nice (INPHYNI), 06200, Nice, France }
\author{Nicolas Noiray}
\email{noirayn@ethz.ch}
\affiliation{CAPS Laboratory, Department of Mechanical and Process Engineering, ETH Zürich, 8092 Zürich, Switzerland}

\date{\today}% It is always \today, today,

\begin{abstract}
Synchronization of self-sustained oscillators under fixed-frequency and amplitude forcing is well understood, but how time-varying forcing mangles phase locking has been much less explored. Theory predicts that slow, deterministic modulation of the drive amplitude or frequency can lead to a peculiar synchronization regime characterized by intermittent locking of the oscillation phase beyond the Arnold-tongue boundaries associated with fixed harmonic forcing. We test these predictions in a controllable aeroacoustic self oscillator, i.e, a whistle, that exhibits a robust limit cycle and is subject to external acoustic forcing with programmable frequency and amplitude modulation. Under both slowly varying frequency or amplitude of the forcing, three regimes are observed: (i) strict synchronization (ii) intermittent synchronization, characterized by alternating phase locking and brief phase slip episodes and (iii) no synchronization, with regular phase slips. Particularly in strict synchronization regime, the phase of the oscillator will follow arbitrary slowly-varying drive phase and under amplitude modulation its amplitude fluctuations are strongly suppressed.
\end{abstract}

                              %display desireds
\maketitle
\section{Introduction}
Phase synchronization, or phase locking, of a self-sustained oscillator under periodic forcing refers to the adjustment of the oscillator’s phase to match that of the external periodic drive \cite{adler2006study,Pikovsky_Rosenblum_Kurths_2001}. It has been studied extensively in mechanics, electronics, chemistry, and biology \cite{Pikovsky_Rosenblum_Kurths_2001,strogatz,strogatz2004sync}. A classic result in time‐independent periodic forcing is the formation of Arnold tongues, tongue-shaped regions in the drive amplitude–frequency plane where $n\!:\!m$ phase locking occurs, \tw{with $n$ and $m$ the numbers of oscillation cycles of the drive and oscillator, respectively}. In the setting of periodic drive, appropriate averaging of higher frequency oscillations will lead to an autonomous dynamical system which governs the stability, bifurcation, and transitions of synchronization regime \cite{strogatz2004sync}. 

Realistic forcing, however, often exhibits variations in frequency and amplitude, making the dynamics intrinsically non–autonomous. A growing number of work tackles these issues for non–autonomous oscillators \cite{jensen2002,Gandhi2015,suprunenko2013chronotaxic,lucas2018stabilization,newman2021stabilization}. \tw{Related rate-dependent phenomena have also been reported in musical-acoustics models, where slowly varying control parameters reshape basins of attraction and alter transient dynamics \cite{bergeot2024,terrien2025basins}}. For a one–dimensional phase oscillator with time–periodic detuning (the non–autonomous Adler equation), Gandhi \textit{et al.}  showed rich canard segments along the thin transition layers between them \cite{Gandhi2015}. In parallel, chronotaxic theory classifies self–sustained non–autonomous oscillators with a moving point attractor on the cycle, as a deterministic, perturbation-resistant time-varying frequency motion \cite{suprunenko2013chronotaxic}. 
More recently, Lucas \textit{et al.} \cite{lucas2018stabilization} showed theoretically that deterministic, slowly varying driving can stabilize oscillations beyond the boundaries of the fixed drive Arnold–tongue, by creating a finite–time \textit{intermittent} regime, where system trajectories alternate between intervals with a stable locked point and intervals with no locked point and yet are stable on average. This effect persists in higher–dimensional models as well \cite{lucas2018stabilization}. Building on these results, a finite–time dynamical–systems framework was proposed \cite{newman2021stabilization,newman2024intermittent}. It treats slow modulation explicitly (slow–fast on a bounded slow–time interval), clarifies when phases track a moving attractor, and gives qualitative criteria for finite–time stability under modulation \cite{newman2021stabilization,newman2024intermittent}.

This work aims at building upon these theoretical predictions, and investigating experimentally dynamic synchronization using a controllable physical oscillator. To that end, we employ a whistle and dynamically force it around its natural self-oscillation frequency. In absence of forcing, above a critical air mass flow, the aeroacoustic system becomes linearly unstable due to a constructive interaction between one of the shear layer eigenmodes and the Helmholtz mode of the whistle, and a limit cycle emerges via a supercritical Hopf bifurcation. Recent analyses and experiments on grazing–flow cavities and compact apertures quantify the feedback mechanism between shear–layer dynamics and the acoustic field, the radiation losses, and the nonlinear saturation that sets the oscillation amplitude \cite{bourquard2021whistling,boujo2018,stoychev2024nonlinear}. Weak external acoustic forcing is applied to drive the aeroacoustic self-oscillation, providing a clean testbed for investigating synchronization under modulation. This combination of robust self–sustained oscillations and precise programmable external driving makes the whistle a natural platform for validating non–autonomous synchronization theory in the laboratory.

Near the Hopf bifurcation, the amplitude of the aeroacoustic mode of the whistle admits a Stuart–Landau oscillator description \cite{crawfordknobloch1991}; under weak forcing from an incident sound field with programmable phase or amplitude, standard phase reduction (e.g., \cite{nakao2016phase}) yields a one–dimensional phase equation with time–varying drive. Taking advantage of a single dominant aeroacoustic mode and a straightforward implementation of slow/fast frequency and amplitude sweeps, we carry out a systematic experimental validation of non-autonomous synchronization and demonstrate intermittently stable regime beyond static Arnold–tongue boundaries, as predicted by Lucas \textit{et\,al.} \cite{lucas2018stabilization}. Furthermore, phase tracking of a moving attractor under sufficiently slow modulation yet arbitrary aperiodic signal \cite{newman2021stabilization,newman2024intermittent} is also tested. 

The paper is organized as follows. Section~\ref{sec:slowflow} describes the aeroacoustic system, presents the derivation of the classical slow-flow equations for the oscillator’s amplitude and phase, and discusses the effects of amplitude and frequency modulation. Section~\ref{sec:Exp} presents experimental results, including instantaneous-frequency analysis and synchronization maps (Arnold tongues) under periodic frequency/amplitude modulation, and responses to aperiodic phase modulation. Section~\ref{sec:conclusion} concludes and provides an outlook.

\section{Asymptotic dynamics under time modulation}\label{sec:slowflow}

\begin{figure}[t]
    \centering
    \includegraphics[width=\linewidth]{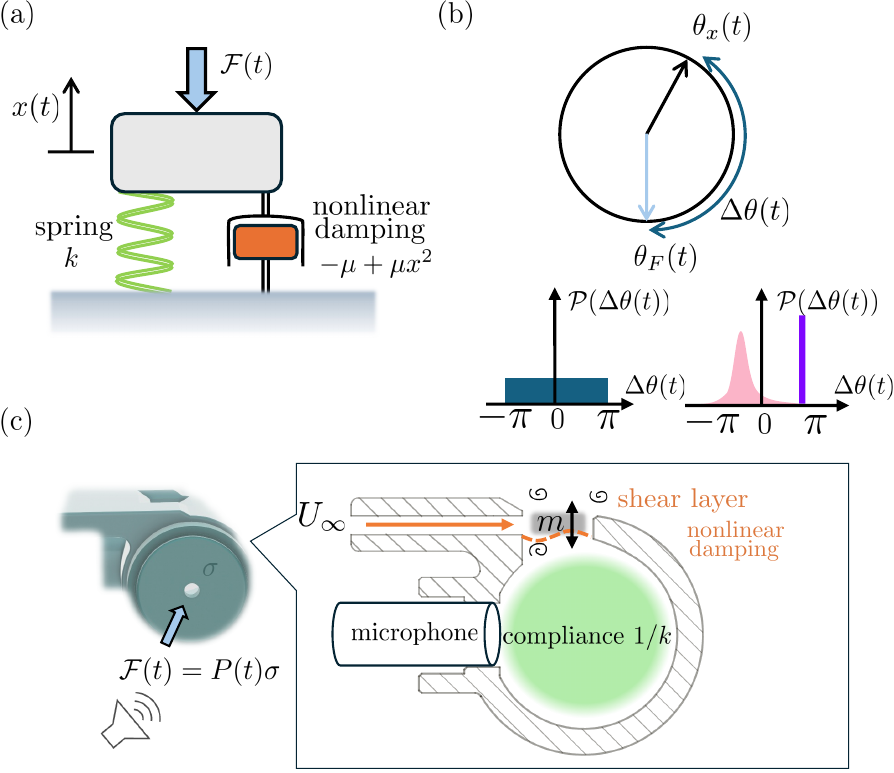}
    \caption{(a) Van der Pol type mass–spring–damper: linear stiffness $k$ (green), negative linear damping coefficient $\mu \in \mathbb{R}^+$ and quadratic nonlinear damping term (orange), and external drive $\mathcal{F}(t)$ (blue).
    (b) Phasor sketch of phase locking: oscillator response (black) and external forcing (blue) on the limit–cycle circle, with examples of probability density functions of the phase difference $\mathcal{P}(\Delta\phi(t))$ in the non-synchronized (blue), quasi-synchronized (pink) and synchronized state (violet).
    (c) Whistle as a self–sustained oscillator: cavity compliance $1/k$ (green), air mass $m$ in the whistle orifice (gray), flow-induced nonlinear negative damping (orange), and side-aperture forcing $\mathcal{F}(t)=P(t)\sigma$ (blue), where the oscillating pressure $P(t)$ is produced by a loudspeaker and $\sigma$ is the aperture area.}
    \label{fig:schematic}
\end{figure}

We model the self-sustained aeroacoustic oscillations in the whistle subject to external acoustic forcing as a Van der Pol (vdP) oscillator weakly forced by an external drive. In Fig.~\ref{fig:schematic}(a), the vdP oscillator is sketched as a mass–spring–damper system, and Fig.~\ref{fig:schematic}(c) shows the aeroacoustic analogue, in which the cavity compliance, the air mass in the whistle orifice, and the shear-layer inducing a nonlinear negative acoustic resistance correspond to the mechanical elements.

In dimensional form, the vdP equation reads 
\begin{equation}
\label{eq:vdP_slow_mod}
  \ddot{p}
  \;-\; \varepsilon\mu\left(1-\frac{p^2}{R^2}\right)\dot{p}
  \;+\; \omega_n^2 p
  \;=\; \varepsilon \mathcal{F}(t),
\end{equation}
where $\mu>0$ defines the linear component of the negative nonlinear damping induced by the air flow, which gives birth to a stable limit cycle for flow rates that exceed the Hopf bifurcation threshold. In equation \eqref{eq:vdP_slow_mod}, $R$ is the limit-cycle amplitude of the unforced vdP oscillator, i.e when $\mathcal{F}(t)=0$, and $\omega_n$ is the natural frequency. The variable $p(t)$ corresponds to the acoustic pressure in the whistle cavity. The small parameter $0<\varepsilon\ll1$ makes explicit that we consider weakly damped, weakly forced self–oscillations, so that the envelope evolves on a slow time scale.

The acoustic forcing imposed by an external loudspeaker through a side aperture is expressed as
\begin{equation}
  \mathcal{F}(t)
  = F\,A(t)\,
  \cos\bigl(\Omega_F(t)\,t + \phi_0\bigr),
\end{equation}
with $F=\mathcal{O}(1)$, $A(t)$ and $\Omega_F(t)$ the slowly varying drive amplitude and frequency, and $\phi_0$ the phase at origin, which, without loss of generality, is set to zero in the following.

In the weak forcing, time independent case ($A\equiv1$, $\Omega_F\equiv\Omega_u=\mathrm{const.}$, with $\Omega_u$ the unmodulated drive frequency), writing the detuning as $\Delta\omega=\Omega_u-\omega_n$ and approximating
\[
  p(t)\approx r\cos(\Omega_u t+\theta),
\]
where $r$ is the forced limit–cycle amplitude, the system exhibits phase synchronization when the phase difference $\theta$ approaches a constant value \cite{Pikovsky_Rosenblum_Kurths_2001,balanov2009}. Equivalently, the phase difference $\Delta\theta(t)=\theta_p(t)-\theta_F(t)$ between oscillator and drive can be viewed as an angle on the unit circle. Thus, synchronization corresponds to $\Delta\theta(t)$ clustering around a fixed value so that its distribution $P(\Delta\theta)$ is sharply peaked rather than uniform, as sketched in Fig.~\ref{fig:schematic}(b). Plotting the synchronization regions in the forcing amplitude and detuning plane yields the classical Arnold tongues \cite{balanov2009,Pikovsky_Rosenblum_Kurths_2001}.

Here we focus on nonautonomous cases where $A(\varepsilon t)$ and/or $\Omega_F(\varepsilon t)$ vary slowly in time. To analyze such modulation, we introduce a standard multiple time scales reduction \cite{nayfeh_nonlinear},
\begin{align}
  &T_0=t,\qquad T_1=\varepsilon t,\qquad 0<\varepsilon\ll1,
  \notag\\
  &p(t)
  \;=\; p_0(T_0,T_1)
  \;+\; \varepsilon\,p_1(T_0,T_1)
  \;+\; \mathcal O(\varepsilon^2),
  \label{eq:multi_expansion}
\end{align}
so that the amplitude and phase evolve on the slow time $T_1$. Subsequently, we will also introduce the explicit modulation phase
\begin{equation}
      \theta_m(T_1)=\varepsilon\Omega_m t = \Omega_m T_1,
\end{equation}
with modulation frequency $\Omega_m$. Thus $\theta_m$ advances on the slow time scale $T_1$, and $\Omega_m$ is chosen much smaller than the natural frequency of the self-oscillator. This classical procedure (e.g., \cite{nayfeh_nonlinear,warminski2001}) yields coupled slow–flow equations for the instantaneous amplitude and phase, capturing synchronization, phase slips, and the effects of slow modulation.

In what follows we treat \textit{amplitude modulation} and \textit{frequency (phase) modulation} separately. For each case we derive the slow-flow equations, identify the locking criteria, and discuss the implications for the experimentally observed synchronization maps.

Regarding the validity of the present analysis, in contrast with the case of fixed amplitude harmonic forcing, the perturbation approximation incurs a $\mathcal{O}(\varepsilon)$ error on time intervals $T=\mathcal{O}(1/\varepsilon)$ in the slow-modulation regime, which is however sufficient for our experimental comparisons.

It is also important to note that Adler equations are often derived for finding locking criteria without accounting for amplitude dynamics, e.g \cite{jensen2002,jensen1998} or \cite{nakao2016phase} for more modern analysis, by assuming that the oscillator amplitude is confined to its limit cycle.

By contrast, we derive the coupled slow-flow equations for both amplitude and phase, which provides a complete picture that is essential for self-oscillators driven by frequency or amplitude modulated forcing.

\subsection{Frequency Modulation}
The case of a slowly varying frequency of the external drive is considered first with the time scales \(T_0=t\) and \(T_1=\varepsilon t\) and \(A\equiv1\). The forcing now reads
\begin{equation}
\label{eq:vdP_fm_main}
\mathcal{F}(T_0,T_1)
  = F\cos \left(\Omega_u T_0+ k_{f}\sin \theta_m(T_1)\,\right),
\end{equation}
with the mean (unmodulated) drive angular frequency $\Omega_u$, frequency–modulation depth \(k_f\), and modulation phase \(\theta_m(T_1)=\Omega_m T_1\).
Using complex notation, we write the forcing as
\[
  \mathcal{F}(T_0,T_1) = \Re\left\{ F\,e^{\,i\Phi(T_0,T_1)}\right\},
\]
and \[  \Phi(T_0,T_1) = \Omega_u T_0 + k_f \sin\theta_m(T_1),\] so that
\begin{equation}
  \mathcal{F}(T_0,T_1)
  = \Re\left\{ F\,e^{\,i\Omega_u T_0}\,e^{\,i k_f \sin\theta_m(T_1)}\right\}.
\end{equation}
The Jacobi–Anger expansion \cite{colton1998} is applied only to the slowly varying factor,
\begin{equation}
  e^{\,i k_f \sin\theta_m(T_1)} = \sum_{n=-\infty}^{\infty}J_n(k_f)\,e^{\,i n\theta_m(T_1)},
\end{equation}
so that the complex forcing becomes
\[
  F\,e^{\,i\Phi(T_0,T_1)}
  = F\sum_{n=-\infty}^{\infty}
      J_n(k_f)\,e^{\,i\bigl[\Omega_u T_0 + n\Omega_mT_1\bigr]}.
\]
Substituting the leading-order terms into the perturbed vdP equation and applying the multiple scales method, only those near resonance forcing terms with carrier $e^{i\omega_n T_0}$ contribute to the slow evolution and all other fast oscillatory terms average out. Keeping the near-resonant terms (here $n=0,\pm1$) then yields the slow–flow system for the envelope amplitude $r(T_1)$ and phase $\theta(T_1)$,
\begin{align}
  \frac{dr}{dT_1}
  =& \frac{\mu}{2}\,r\left(1 - \frac{r^2}{R^2}\right)
     - \frac{F\,J_0(k_f)}{2\,\omega_n}\,\sin\theta\notag\\
     &- \frac{F\,J_1(k_f)}{\omega_n}\,\sin\bigl(\theta-\theta_m\bigr),
  \label{eq:slow_amp_FM}\\
  \frac{d\theta}{dT_1}
  = &\Delta\omega
     - \frac{F\,J_0(k_f)}{2\,\omega_n\,r}\,\cos\theta
     - \frac{F\,J_1(k_f)}{\omega_n\,r}\,\cos\bigl(\theta-\theta_m\bigr),
  \label{eq:slow_phase_FM}
\end{align}
where \(\Delta\omega=\Omega_u-\omega_n\) is the detuning. Taking \(k_f=0\) (so \(J_0(k_f)=1\), \(J_1(k_f)=0\)) reduces \eqref{eq:slow_amp_FM}–\eqref{eq:slow_phase_FM} to the standard truncated slow–flow equations of the harmonically forced vdP oscillator \cite{balanov2009}.

When deriving equations \eqref{eq:slow_amp_FM}–\eqref{eq:slow_phase_FM}, the truncated terms are only to the $n=0,\pm1$ from the Jacobi–Anger expansion, i.e.\ the $J_0$ and $J_1$ terms. For small $k_f$, $J_n(k_f)=\mathcal O \big((k_f/2)^{|n|}\big)$ so $|J_{\pm2}|=\mathcal O(k_f^2)$, and the associated terms are both weak and off–resonant, thus they average out at leading order. The resulting slow flow is accurate up to $\mathcal O(k_f^2)$ corrections. Including $J_2$ would add terms proportional to $\cos(\theta-2\theta_m)$ and $\sin(\theta-2\theta_m)$ with coefficients proportional to $J_2(k_f)$.

\begin{figure}[t]
    \centering
    \includegraphics[width=0.8\linewidth]{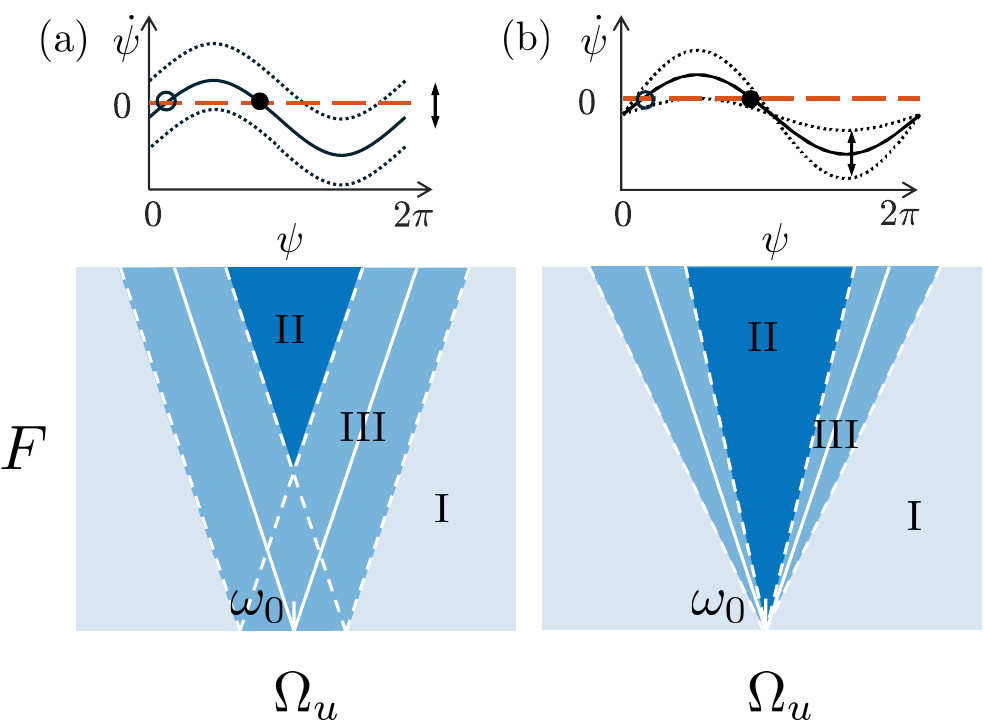}
    \caption{Schematic view of synchronization diagram under slow modulation. Top: time evolution over one modulation cycle of the phase dynamics (solid black), together with horizontal lines (orange dashed) indicating when fixed points exist (stable fixed points are marked as filled circles while unstable ones marked as open circles). Bottom: corresponding schematic Arnold tongues in the plane of averaged forcing frequency (horizontal) and forcing amplitude (vertical), showing three regimes: light blue—no synchronization (region I), medium blue—intermittent synchronization (region III), and dark blue—persistent synchronization (region II). \tw{The solid white boundary is the classical (unmodulated) locking threshold. The white dashed boundaries indicate the locking thresholds evaluated at the two extrema of the modulation.} (a) Frequency modulation: the forcing frequency oscillates in time while the forcing amplitude is fixed. Panel is redrawn, with notation adapted from Lucas \textit{et\,al.}~\cite{lucas2018stabilization}. (b) Amplitude modulation: the forcing amplitude oscillates while the forcing frequency is fixed. \tw{The inner white dashed lines correspond to the persistent synchronization threshold given by Eq.~\eqref{eq:AM_threshold_persistent}, and the outer white dashed lines correspond to the broad synchronization threshold given by Eq.~\eqref{eq:AM_threshold_broad}}}
    \label{fig:stability_scheme}
\end{figure}

For a phase oscillator with slowly varying detuning, Lucas \textit{et\,al.}~\cite{lucas2018stabilization} studied the nonautonomous Adler equation \(\dot\psi=\Delta\omega(t)+\gamma\sin\psi\), where \(\psi\) is the phase difference, \(\Delta\omega(t)\) a prescribed time-dependent frequency detuning, and \(\gamma>0\) the coupling constant. Instantaneous fixed points satisfy \(\Delta\omega(t)+\gamma\sin\psi^*(t)=0\) and exist only when \(|\Delta\omega(t)|\le\gamma\). According to how often this inequality holds over one modulation period, they distinguished three regimes (I–III in their Fig.\,1 and also in Fig.~\ref{fig:stability_scheme}): no synchronization (\(|\Delta\omega(t)|>\gamma,\,\forall t\)), persistent synchronization (\(|\Delta\omega(t)|<\gamma,\,\forall t\)), and an \emph{intermittent} regime in which \(|\Delta\omega(t)|\) crosses \(\gamma\) twice per cycle, where fixed points are created and destroyed twice per cycle. In the intermittent case the dynamics are deterministic. The system alternates between intervals of convergence toward the instantaneous fixed point and intervals of drift, producing one slip per half-cycle of the slow modulation. Lucas \textit{et\,al.} showed that, in the plane of mean detuning versus coupling, the union of the persistent and intermittent regimes occupies a larger region than the static tongue \(|\Delta\omega|<\gamma\). The apparent widening of the Arnold tongue under modulation is due to the intermittent regime, and parameter values with negative largest Lyapunov exponent of \(\psi\) are regarded as synchronized in a broad sense \cite{lucas2018stabilization}.

In the present work we adopt the same nonautonomous modulation, but start from a full van der Pol model: we first derive slow–flow equations for the amplitude and phase of the forced vdP oscillator and then show (Appendix~\ref{sec:appendix}) that the slow phase difference reduces, after an explicit change of variables, to a nonautonomous Adler equation of the form studied by Lucas \textit{et\,al.}~\cite{lucas2018stabilization}. 

For each frozen modulation phase \(\theta_m\), instantaneous equilibria of the slow–flow phase equation \eqref{eq:slow_phase_FM} satisfy
\[
  0
  = \Delta\omega
    - \frac{F\,J_0(k_f)}{2\,\omega_n\,r}\,\cos\theta^*
    - \frac{F\,J_1(k_f)}{\omega_n\,r}\,\cos\bigl(\theta^*-\theta_m\bigr),
\]
which defines a time-dependent locking condition. As detailed in Appendix~\ref{sec:appendix}, a change of variables from \(\theta\) to the phase difference \(\psi\) recasts the slow phase dynamics into a nonautonomous Adler equation
\begin{equation}
  \frac{d\psi}{dT_1}
  = \Delta\omega_{\mathrm{eff}}(T_1)
    + \gamma_{\mathrm{eff}}(T_1)\,\sin\psi,
  \label{eq:FM_adler_main}
\end{equation}
with a time-varying effective detuning \(\Delta\omega_{\mathrm{eff}}(T_1)\) and coupling \(\gamma_{\mathrm{eff}}(T_1)\) that are explicit functions of \(\theta_m(T_1)=\Omega_m T_1\). For frozen \(T_1\), instantaneous fixed points of \eqref{eq:FM_adler_main} exist if and only if \(|\Delta\omega_{\mathrm{eff}}(T_1)|\le \gamma_{\mathrm{eff}}(T_1)\), exactly as in the static Adler case.

In the small–modulation regime \(k_f\ll 1\) one has \(J_0(k_f)\approx 1\), \(J_1(k_f)\approx k_f/2\), and the coupling is nearly constant while the detuning oscillates. In leading order,
\begin{equation}
  \gamma_{\mathrm{eff}}=\frac{F}{2\,\omega_n\,r},
  \quad
  \Delta\omega_{\mathrm{eff}}(T_1)
  = \Delta\omega - \Omega_m k_f\cos\theta_m(T_1),
\end{equation}
so that \eqref{eq:FM_adler_main} reduces to the form analyzed by Lucas \textit{et\,al.}~\cite{lucas2018stabilization}. Using their classification, three regimes follow from how often the instantaneous inequality \(|\Delta\omega_{\mathrm{eff}}(T_1)|\le\gamma_{\mathrm{eff}}\) is satisfied over one modulation period: 
\begin{align}
\text{No synchronization:}\quad
& \gamma_{\mathrm{eff}} < |\Delta\omega| - \Omega_m k_f, \label{eq:FM_regionI_main}\\
\text{Persistent synchronization:}\quad
& \gamma_{\mathrm{eff}} \ge |\Delta\omega| + \Omega_m k_f, \label{eq:FM_regionII_main}\\
\text{Intermittent synchronization:}\notag\\
 |\Delta\omega| - \Omega_m k_f \le &\gamma_{\mathrm{eff}} \le |\Delta\omega| + \Omega_m k_f. 
\label{eq:FM_regionIII_main}
\end{align}
In the intermittent regime, \(|\Delta\omega_{\mathrm{eff}}(T_1)|=\gamma\) is crossed twice per modulation period, two saddle–node bifurcations occur, and the trajectory alternates deterministically between synchronized and nonsynchronized subintervals. These three regions correspond to the schematic phase diagram shown in Fig.~\ref{fig:stability_scheme}(a) and to regions I–III in Ref.~\cite{lucas2018stabilization}.
In the experimental and schematic Arnold–tongue plots (Fig.~\ref{fig:stability_scheme}) we use the plane of mean drive frequency and forcing amplitude $(\Omega_u,F)$, which is directly related to the theoretical geometry in $(\Delta\omega,\gamma)$ as discussed in Ref.~\cite{lucas2018stabilization}.

A complementary, heuristic view is treating \eqref{eq:slow_phase_FM} as a static forcing term plus a near-resonant sideband generated by frequency modulation. The two contributions combine into an effective forcing scale
\begin{equation}
  F_{\mathrm{eff}}(k_f)
  = \frac{F}{2\,\omega_n\,r}\,
    \sqrt{\,J_0(k_f)^2 + \left[2\,J_1(k_f)\right]^2\,},
\end{equation}
where the square root reflects the root-mean-square combination of the carrier and first sideband amplitudes. 

A rough estimate of the broadened locking interval uses the static criterion \(|\Delta\omega|\lesssim F_{\mathrm{eff}}\) applied quasi-statically over a modulation period. For small \(k_f\) one has \(J_0(k_f)\approx 1\) and \(J_1(k_f)\approx k_f/2\), giving $F_{\mathrm{eff}}\approx F\sqrt{ 1+k_f^2\,}/(2 \omega_n r).$ Thus, frequency modulation redistributes drive power between the carrier and its sidebands, slightly increasing the effective coupling and widening the locking range in the broad sense of synchronization without changing the nominal drive amplitude.

\subsection{Amplitude modulation}
The case of amplitude modulation of the harmonic forcing is now considered. The amplitude modulation effects the instantaneous coupling strength and therefore the locking condition in time, producing the same three regimes observed under frequency modulation (persistent, intermittent and no synchronization regimes), but with a different alternation of the Arnold tongue. The forcing is now taken as
\begin{equation}
  \mathcal{F}(T_0,T_1)
  = F\left[1+k_a\sin\theta_m(T_1)\right]\cos(\Omega_u T_0),
  \label{eq:vdP_am_main}
\end{equation}
with now constant drive frequency \(\Omega_u\), amplitude modulation depth \(k_a\), and modulation phase \(\theta_m(T_1)=\Omega_m T_1\).

With the slow time $T_1$ defined above and the modulation phase $\theta_m$ already introduced, the slow-flow for the coupled amplitude–phase dynamics is now
\begin{align}
  \frac{d r}{d T_1}
    &= \frac{\mu}{2}\,r\Bigl(1 - \frac{r^2}{R^2}\Bigr)
      - \frac{F}{2\,\omega_n}
        \bigl(1 + k_{a}\sin \theta_m\bigr)\,
        \sin\theta,
  \label{eq:slow_amp_AM}\\[2pt]
  \frac{d \theta}{d T_1}
    &= \Delta \omega
      - \frac{F}{2\,\omega_n\,r}
        \bigl(1 + k_a\sin \theta_m\bigr)\,
        \cos\theta,
  \label{eq:slow_phase_AM}
\end{align}
where \(0<k_a<1\) is the input amplitude modulation strength and $F$ is the nominal forcing level.

We first analyze persistent synchronization and the attenuation of the imposed modulation. From the phase slow-flow \eqref{eq:slow_phase_AM} it is convenient to define the time–dependent effective coupling
\begin{equation}
  \gamma_{\mathrm{eff}}(T_1)=\frac{F}{2\,\omega_n\,r(T_1)}\,
  \left[1+k_a\sin\theta_m(T_1)\right].
  \label{eq:AM_gamma_eff}
\end{equation}
For frozen $T_1$, equilibria of \eqref{eq:slow_phase_AM} exist if and only if
\begin{equation}
  |\Delta\omega|\le |\gamma_{\mathrm{eff}}(T_1)|.
  \label{eq:AM_inst_cond}
\end{equation}
Thus persistent synchronization throughout the modulation requires the uniform condition
\begin{equation}
  |\Delta\omega|\le\min_{T_1} |\gamma_{\mathrm{eff}}(T_1)|.
  \label{eq:AM_persistent_cond}
\end{equation}
Inside the synchronization regime the phase $\theta(T_1)$ remains close to a slowly varying locked value, which we denote by $\theta_s(T_1)$, and the amplitude remains near a forced limit–cycle value $R_F$. Writing
\[
  r(T_1) = R_F + \delta r(T_1),\quad |\delta r|\ll R_F,
\]
and linearizing \eqref{eq:slow_amp_AM} near $R_F$ gives a sinusoidal response of the amplitude at the modulation frequency. If we write this response as
\[
  r(T_1) = R_F\left(1+k_p\sin\theta_m(T_1)\right),
\]
where $k_p$ is the normalized output modulation depth of $p$, we obtain
\begin{equation}
  k_p = \frac{\delta r}{R_F}
  = \frac{F\,\sin\theta_s}{2\,\omega_n\,R_F\,\mu}\,k_a
  = \Bigl(1-\frac{R}{R_F}\Bigr)\,k_a.
  \label{eq:k_in_out}
\end{equation}
Since $R_F>R$, the factor $1-R/R_F$ is strictly between 0 and 1. This implies that slow amplitude modulation is attenuated ($k_p<k_a$). To leading order in $k_a$, it is therefore consistent to treat the amplitude as quasi-constant and replace $r(T_1)$ by $R_F$ in the coupling.

With this approximation the effective coupling becomes
\begin{equation}
  \gamma_{\mathrm{eff}}(T_1) \approx  \gamma_0\left[1+k_a\sin\theta_m(T_1)\right],
  \quad\gamma_0= \frac{F}{2\,\omega_n\,R_F}.
  \label{eq:AM_gamma_approx}
\end{equation}
The minimum and maximum values over a modulation period are
\[
  \gamma_{\min}=\gamma_0(1-k_a),\quad \gamma_{\max}=\gamma_0(1+k_a).
\]
If the detuning lies in the window
\begin{equation}
    \gamma_{\min}<|\Delta\omega|<\gamma_{\max},
    \label{eq:AM_window}
\end{equation}
then the similar intermittent synchronization would occur. The crossing times satisfy
\begin{equation}
  \sin\theta_m(T_1)= \frac{|\Delta\omega|/\gamma_0 - 1}{k_a}\in[-1,1],
  \label{eq:AM_crossing}
\end{equation}
so there are two saddle–node bifurcations per modulation cycle in the intermittent synchronization regime.

Moreover, the straight lines \(|\Delta\omega|= \gamma_0(1\pm k_a)\) divide the geometry of synchronization diagram in the plane of detuning and nominal coupling $(\Delta\omega,\gamma_0)$ into regions I-III as introduced before. The persistent synchronization region satisfies
\[
  |\Delta\omega|\le\gamma_0(1-k_a),
\]
the intermittent synchronization region has the two-sided collar
\[
  \gamma_0(1-k_a)
  < |\Delta\omega| <
  \gamma_0(1+k_a),
\]
and no synchronization occurs when
\[
  |\Delta\omega|\ge\gamma_0(1+k_a).
\]
Thus persistent synchronization forms an inner triangle in the $(\Delta\omega,\gamma_0)$ plane, while broad synchronization (persistent plus intermittent) occupies a larger outer triangle, with the intermittent regime filling the triangular side bands between the two boundaries which are marked as white dashed lines in Fig.~\ref{fig:stability_scheme}.

At fixed detuning $|\Delta\omega|$, these conditions can be rewritten in terms of the forcing amplitude $F$. Any broad synchronization requires
\begin{equation}
  F \ge \frac{2\,\omega_n\,R_F}{1+k_a}\,|\Delta\omega|,
  \label{eq:AM_threshold_broad}
\end{equation}
while persistent synchronization requires
\begin{equation}
  F \ge \frac{2\,\omega_n\,R_F}{1-k_a}\,|\Delta\omega|.
  \label{eq:AM_threshold_persistent}
\end{equation}
Amplitude modulation therefore lowers the threshold for synchronization by the factor \(1/(1+k_a)\) in the broad sense (persistent plus intermittent) but raises the threshold for persistence to \(1/(1-k_a)\), producing the widened outer triangle and the shrunken persistent triangle sketched in Fig.~\ref{fig:stability_scheme}(b).

To compare the effects of different slow modulation, frequency modulation acts as a time-varying detuning, thus moving along the classical boundary corresponds to shifting left and right in the $(\Delta\omega,\gamma)$ plane. The broad synchronization region therefore becomes a trapezoid symmetric about the static Arnold–tongue boundary. In contrast, amplitude modulation acts as a time-varying coupling, so moving along the classical boundary corresponds to shifting up and down in $(\Delta\omega,\gamma)$. The intermittent band then appears as two triangular side regions around a narrower persistent triangle, and the overall broad Arnold tongue widens accordingly.

\section{Experimental Results and Discussion}\label{sec:Exp}
Figure~\ref{fig:schematic}(c) illustrates our experimental externally driven self-oscillator. It is a 3D printed whistle following the design presented in \cite{stoychev2025synchronization}, that is mounted as an obstacle in the middle of a one-dimensional waveguide. Air is supplied to the whistle via a volume flow controller (Bronkhorst). The air mass flow is adjusted to induce self-oscillations, while acoustic coupling to the waveguide is achieved through 4~mm apertures on the two lateral sides of the whistle. \tw{In such ducted configurations, mean-flow and acoustic interaction can be viewed as providing an effective source of acoustic gain and loss \cite{auregan2017}}. The forcing with incident acoustic waves is generated by a compression driver (CP850Nd, Beyma) located at one end of the waveguide. The waveguide, whistle, and external forcing arrangement are essentially the same as in Ref.~\cite{stoychev2025synchronization}, where the focus was on acoustic wave scattering under steady constant driving, whereas here the same platform is used to study synchronization under slowly varying, nonstationary forcing. In the experiments a dimensionless drive amplitude $F$ is prescribed in the control script, converted to a voltage command for the power amplifier and compression driver. The pressure inside the whistle is recorded using a microphone (46BD-FV G.R.A.S.), which is flush-mounted on the internal wall of the whistle. In the following analysis, an air flow of 10 L/min is steadily supplied to the whistle, resulting in self-sustained oscillations at a natural whistling frequency of 1160 Hz.

To quantify the degree of synchronization between the whistle and the external drive, the Phase Locking Value (PLV) employed in \cite{lachaux1999,aydore2013} is used. Alternative metrics (such as the largest Lyapunov exponent estimated from delay embeddings \cite{rosenstein1993} or phase–difference entropy \cite{tass1998detection}) turned out to be less robust in the present noisy, single–observable system. The instantaneous phases, $\theta_p(t_k)$ and $\theta_F(t_k)$, are obtained from taking the Hilbert transform of pressure signal $p(t)$ and force $\mathcal{F}(t)$ at sampling times $t_k$. The raw phase difference sequence is then
\begin{equation}
  \Delta\theta_k = \theta_p(t_k) - \theta_F(t_k).
\end{equation}
With slow modulation of the forcing (in frequency or amplitude), $\Delta\theta_k$ contains a deterministic component at the modulation phase $\theta_m(t)$. To isolate the residual clustering, the phase difference is demodulated by projecting $\Delta\theta_k$ onto the known drive waveform and subtracting the fitted component. Specifically, $\Delta\theta_k$ is approximated as \begin{equation}
  \Delta\theta_k
  = a\,\theta_F(t_k) + \Delta\theta_k^{\perp},
  \label{eq:phase_fit}
\end{equation} where
\[a = \frac{\sum_{k=1}^{N} \theta_F(t_k)\,\Delta\theta_k}
     {\sum_{k=1}^{N} \theta_F(t_k)^{2}},
\] and $\Delta\theta_k^{\perp}$ is the modulation-corrected phase. This projection removes the known deterministic drift from $\Delta\theta_k$. The modulation-corrected PLV (mcPLV) is then defined as
\begin{equation}
  \mathrm{mcPLV} = \left|\frac{1}{N}\sum_{k=1}^{N}
    e^{\,i\,\Delta\theta^{\perp}_k}\right|,
  \quad 0 \le \mathrm{mcPLV} \le 1.
  \label{eq:mcPLV}
\end{equation}
Values near 1 indicate persistent synchronization, intermediate values typically reflect intermittent synchronization and values near 0 indicate no synchronization. In the maps below, mcPLV is reported together with circular histograms of $\Delta\theta_k^{\perp}$ for interpretability.

\subsection{Experimental synchronization under time‐periodic modulation of forcing frequency}

Guided by the schematic in Fig.~\ref{fig:stability_scheme}(a), the drive frequency is modulated slowly and synchronization is quantified using the mcPLV defined above. In the frequency-modulation experiments the instantaneous drive frequency is of the form
\[
  \Omega_F(t)
  = \Omega_u + k_f\,\Omega_m\cos(\Omega_m t),
\]
with mean drive frequency $\Omega_u$, dimensionless modulation depth $k_f$, and modulation frequency $\Omega_m$ (here $f_m=\Omega_m/2\pi=1$~Hz). The modulation is therefore much slower than the natural frequency $1160$~Hz of the whistle, so the oscillator experiences a slowly varying detuning on the acoustic timescale. For convenience, the modulation depth is the frequency amplitude
\[
  \Delta f = \frac{k_f\,\Omega_m}{2\pi},
\]
so that $\Delta f$ is the peak deviation of $\Omega_F(t)$ expressed in Hz. 

\begin{figure*}
    \centering
    \includegraphics[width=0.95\linewidth]{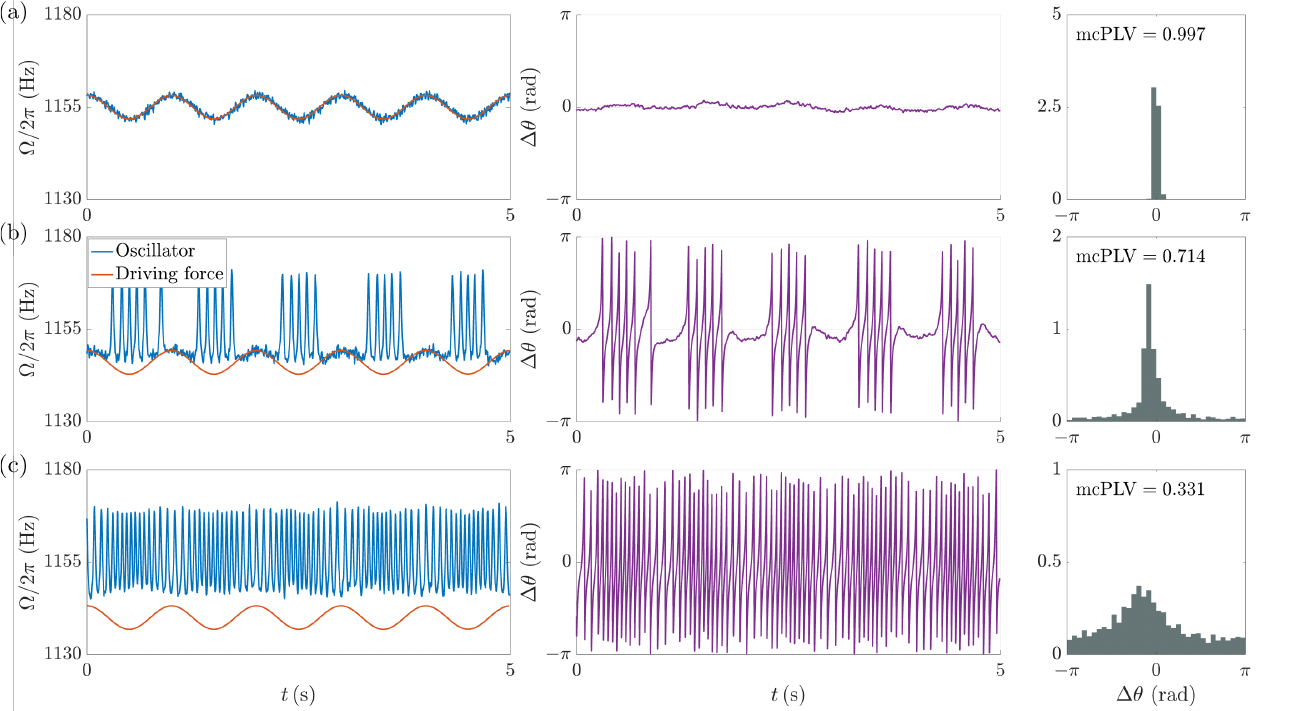}
    \caption{Instantaneous frequency and phase–difference analysis for the three synchronization regimes (I–III) marked in Fig.\ref{fig:freq-mod}(c), at fixed drive level $F=1.85$ and mean drive frequencies $\Omega_u/2\pi = 1155,\,1146,\,1140$~Hz for (a), (b), and (c), respectively. Each row corresponds to one regime: left, instantaneous frequencies of the drive (orange) and oscillator (blue); middle, wrapped modulation-corrected phase difference $\Delta\theta(t)$; right, probability density of $\Delta\theta$ together with the corresponding mcPLV. Panel (a) illustrates persistent synchronization (regime II), (b) intermittent synchronization with periodic phase slips (regime III), and panel (c) no synchronization (regime I). Tighter clustering and thus mcPLV value closer to 1 indicates stronger long–term synchronization.}
    \label{fig:inst_phase}
\end{figure*}

\begin{figure}
    \centering
    \includegraphics[width=1\linewidth]{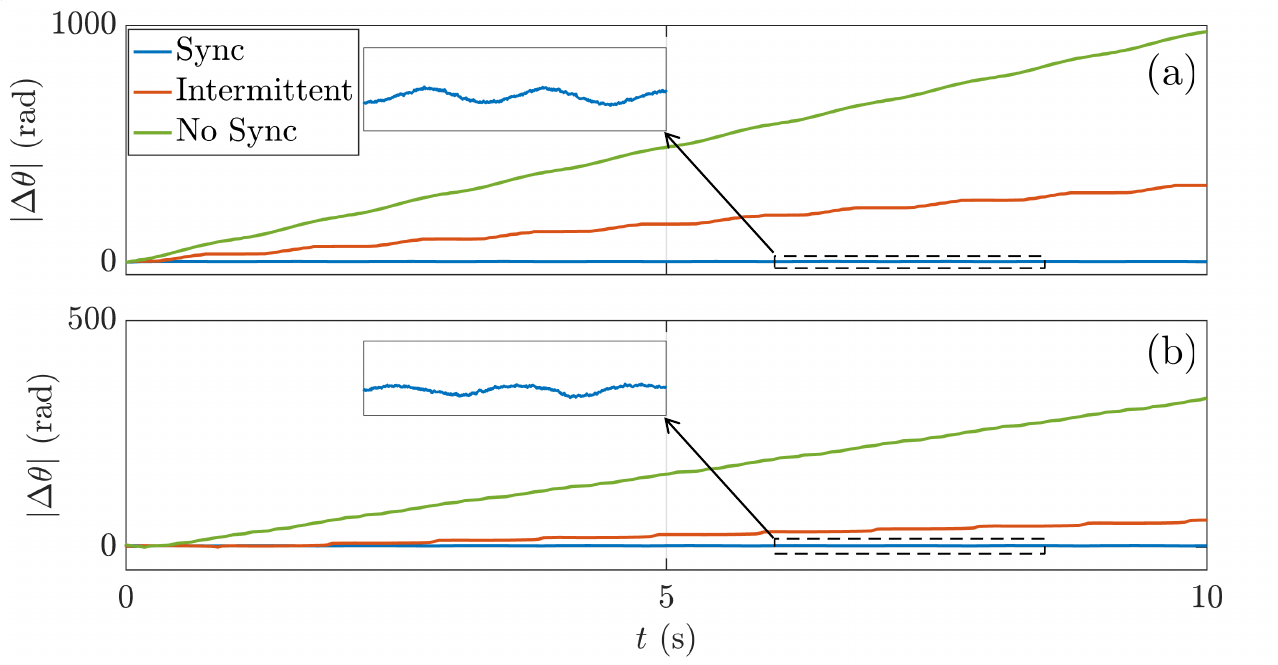}
    \caption{Phase slip dynamics for the three regimes. (a) amplitude modulation; (b) frequency modulation. Each panel plots the absolute value of the unwrapped phase difference, $|\Delta\theta(t)|$, for representative points. Colors: blue—persistent synchronization (bounded, no steps), red—intermittent synchronization (plateaus separated by discrete phase slips), green—no synchronization (diffusive drift with no plateau structure).}
\label{fig:phase-slip}
\end{figure}

Figure~\ref{fig:inst_phase} shows representative time series from three points in the synchronization map corresponding to persistent synchronization, intermittent synchronization, and no synchronization. In the persistent regime, the instantaneous response frequency $\omega_p(t)$ (blue) closely follows the modulated drive $\Omega_F(t)$ (orange), the unwrapped phase difference $\Delta\theta(t)$ remains confined near a constant offset (see Fig.~\ref{fig:phase-slip}), its circular histogram is sharply peaked, and the mcPLV is close to 1. In the intermittent regime, each epoch of entrainment is interrupted by phase slips. During synchronized subintervals $\omega_p(t)\approx\Omega_F(t)$, slips appear as stepwise jumps in $\Delta\theta(t)$, producing multiple peaks in the histogram and intermediate mcPLV values between 0 and 1. The slip timing is set by the slow modulation, with one slip per modulation cycle, consistent with two saddle-node crossings per period when the instantaneous detuning traverses the static tongue boundary in Fig.~\ref{fig:stability_scheme}(a) and with the nonautonomous Adler dynamics described by Lucas \textit{et\,al.}~\cite{lucas2018stabilization}. Qualitative phase-slip patterns are summarized in Fig.~\ref{fig:phase-slip}(b), where persistent traces remain bounded, intermittent traces exhibit one phase slip per modulation cycle, and no synchronized traces grow without bound.

\begin{figure}
    \centering
    \includegraphics[width=1\linewidth]{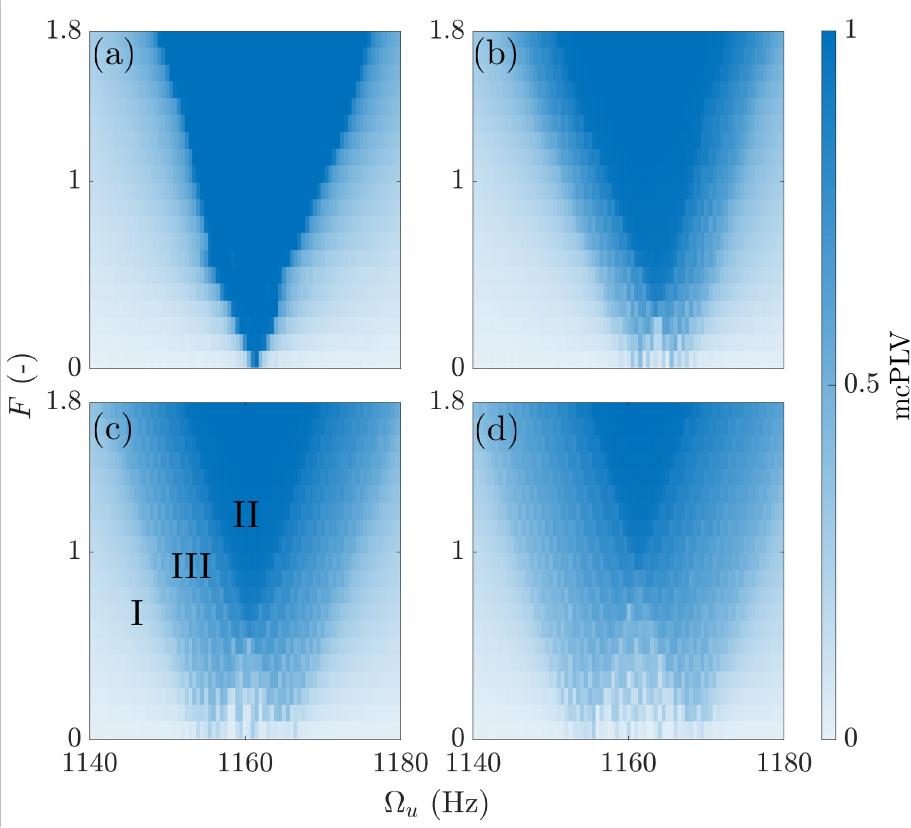}
    \caption{Experimentally obtained Arnold–tongue maps based on mcPLV under increasing frequency modulation depth. Panels (a–d) show synchronization in the detuning–forcing plane for peak frequency shifts \(\Delta f= 0,\,10/2\pi,\,30/2\pi,\,50/2\pi\) Hz, respectively (color code: light blue—no synchronization, region III; medium blue—intermittent, region II; deep blue—persistent, region I). As \(\Delta f\) increases, the broad synchronized region (persistent and intermittent) widens, with a shrinking persistent core and a growing intermittent band, consistent with the nonautonomous detuning picture. For reference, panel (c) labels regions I–III.}
    \label{fig:freq-mod}
\end{figure}

In the no synchronization regime, $\omega_p(t)$ drifts relative to $\Omega_F(t)$, $\Delta\theta(t)$ behaves as an effectively unbounded walk, the phase-difference histogram is nearly uniform on $(-\pi,\pi]$, and the mcPLV is low. Figure~\ref{fig:freq-mod} assembles mcPLV-based Arnold tongue diagrams in the $(\Omega_u,F)$ plane for increasing modulation depth ($\Delta f = 0,\,10/2\pi,\,30/2\pi, 50/2\pi$~Hz). Here $F$ denotes the same dimensionless drive level used in the control script. For $\Delta f=0$ the classical V-shaped static tongue is recovered. As the modulation depth increases, the broad synchronization region (persistent plus intermittent synchronization) expands such that the boundary between no synchronization and intermittent synchronization moves outward, while the boundary between intermittent and persistent synchronization moves inward. Consequently, the persistent synchronization region shrinks, the intermittent synchronization region widens, and their union grows, in qualitative agreement with the regime geometry predicted by the nonautonomous Adler analysis of Lucas \textit{et\,al.}~\cite{lucas2018stabilization}. All together, the instantaneous traces, phase-slip statistics, and mcPLV maps provide an experimental counterpart to the theoretical and numerical picture developed in Ref.~\cite{lucas2018stabilization}.

\subsection{Time-periodic Modulation of forcing Amplitude}

\begin{figure}
    \centering
    \includegraphics[width=0.82\linewidth]{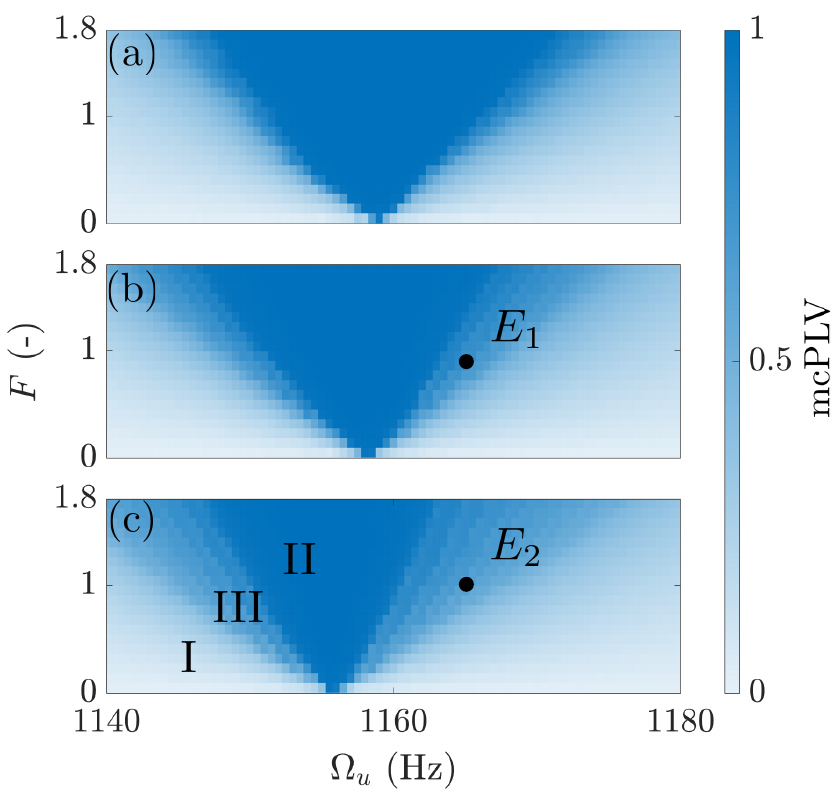}
    \caption{(a–c) show mcPLV-based Arnold tongues in the $(\Omega_u,F)$ plane for increasing amplitude-modulation depth $k_a$. Regions I, II, and III denote no synchronization, persistent synchronization, and intermittent synchronization, respectively (as in Fig.~\ref{fig:stability_scheme}). Example points $E_1$ and $E_2$ in panels (b) and (c) mark two operating conditions inside region III at the same $(\Omega_u,F)$ but different $k_a$ and their time traces are shown in Fig.~\ref{fig:amp_mod_phase}.}
    \label{fig:amp_mod}
\end{figure}

The effect of slow amplitude modulation of the drive on synchronization is now examined, guided by the schematic geometry in Fig.~\ref{fig:stability_scheme}(b). The drive amplitude is modulated slowly according to
\[
  \mathcal{F}(t)
  = F\bigl[1+k_a\sin(\Omega_m t)\bigr]\cos(\Omega_u t),
\]
with mean drive frequency $\Omega_u$, amplitude-modulation depth $k_a$, and modulation angular frequency $\Omega_m$ (again $f_m=\Omega_m/2\pi=1$~Hz, much slower than the natural frequency). Synchronization is quantified using the same mcPLV as in the frequency-modulation case. Figure~\ref{fig:amp_mod} shows mcPLV-based Arnold tongue diagrams in the $(\Omega_u,F)$ plane for increasing modulation depths $k_a=0.15,\,0.30,$ and $0.50$. As predicted by the time-varying coupling phase equation \eqref{eq:slow_phase_AM} and the condition $|\Delta\omega|\le|\gamma_{\mathrm{eff}}(T_1)|$, the persistent synchronization region collapses onto an inner triangle, the intermittent synchronization region forms two triangular collars around it, and the union (broad synchronization) occupies a larger outer triangle whose size increases with $k_a$, as sketched in Fig.~\ref{fig:stability_scheme}(b).

\begin{figure*}
    \centering
    \includegraphics[width=0.95\linewidth]{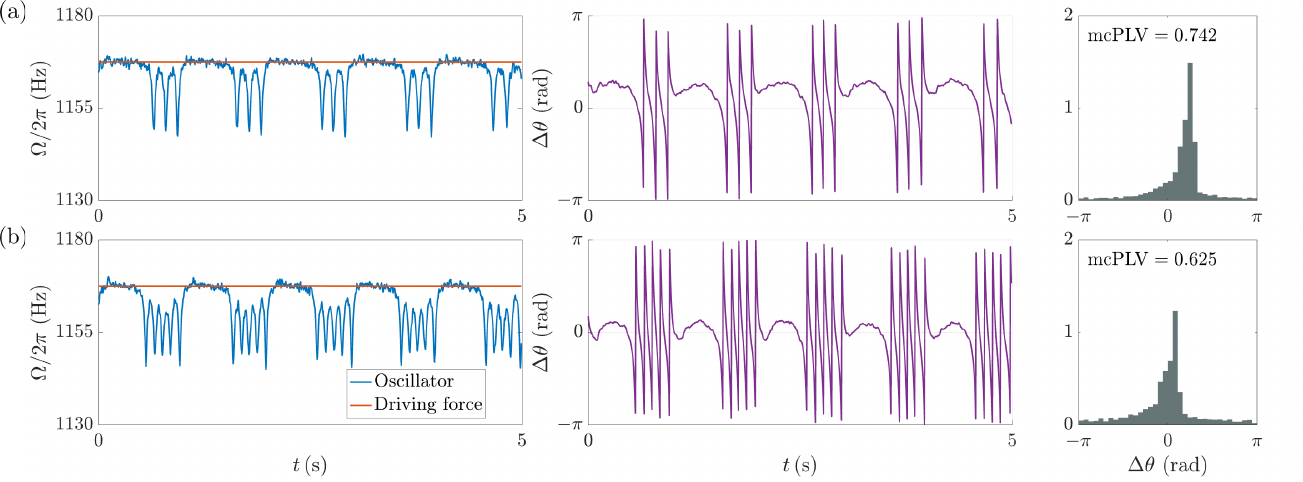}
    \caption{Instantaneous frequencies and phase difference for two intermittent cases $E_1$ and $E_2$ (marked in Fig.~\ref{fig:amp_mod}(b,c)). Both points lie at the same frequency and forcing amplitude but have different amplitude modulation depths $k_a$. As $k_a$ increases from $E_1$ to $E_2$, the fraction of time spent out of synchronization within one modulation period grows and additional phase slips appear in the phase difference, in direct analogy with the intermittent behavior observed under frequency modulation in Fig.~\ref{fig:inst_phase}(b).}
    \label{fig:amp_mod_phase}
\end{figure*}

Time traces at two representative points in the intermittent region ($E_1$: $k_a=0.30$; $E_2$: $k_a=0.50$, with the same detuning and mean forcing level) are shown in Fig.~\ref{fig:amp_mod_phase}. During synchronized subintervals within each modulation cycle, the instantaneous frequencies $\omega_p(t)$ and $\Omega_F(t)$ coevolve and the locked phase offset varies smoothly with the envelope. During nonsynchronized gaps the phase accumulates slips, leading to stepwise jumps in the unwrapped phase difference $\Delta\theta(t)$. For the same operating point in the intermittent region, increasing $k_a$ lengthens the unlocked portions and increases the amount of phase slips per modulation period, consistent with the theoretical picture of two saddle-node crossings of the locking condition per cycle and with the analytical intermittent window. This trend is visible in Fig.~\ref{fig:phase-slip}(a): persistent traces remain bounded, intermittent traces show periodic slips and no synchronization traces show an unbounded drift.

The locked fraction over one slow period can be expressed in terms of the model parameters. Within the intermittent band \eqref{eq:AM_window}, the theory predicts a locking duty cycle
\[
  D_{\mathrm{lock}}
  = \frac{1}{\pi}\arccos \left(\frac{|\Delta\omega|/\gamma_0 - 1}{k_a}\right),
\]
with $D_{\mathrm{lock}}=1$ in the persistent region and $D_{\mathrm{lock}}=0$ above the outer boundary. At a fixed operating point inside the intermittent band, this implies that increasing $k_a$ reduces $D_{\mathrm{lock}}$ (more slip time) when $|\Delta\omega|<\gamma_0$ and increases $D_{\mathrm{lock}}$ when $|\Delta\omega|>\gamma_0$, in qualitative agreement with the measured fraction of time spent in synchronized intervals.

\begin{figure}
    \centering
    \includegraphics[width=1\linewidth]{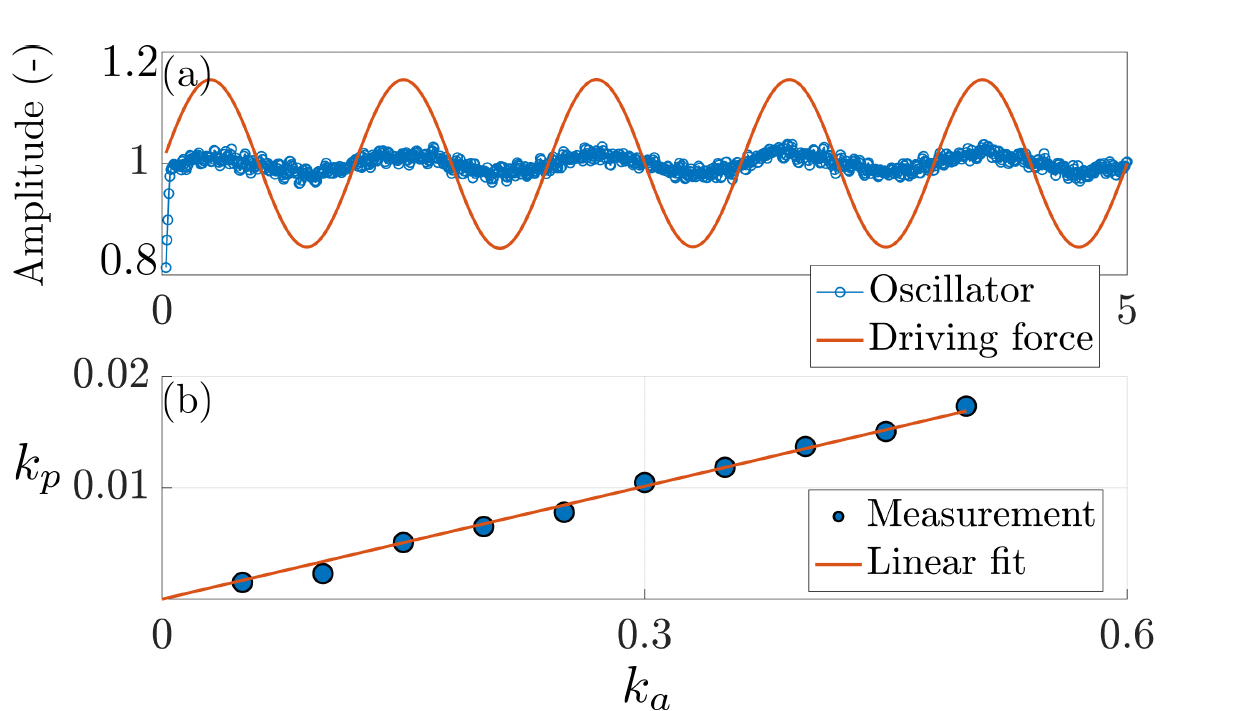}
    \caption{Amplitude response under slow amplitude modulation. Top: normalized amplitude envelopes of drive and oscillator \tw{for $k_a=0.15$}, showing attenuation of the imposed envelope variation in the whistle response. \tw{The output modulation depth is $k_p=0.0051$, obtained by fitting the normalized response envelope with a single-harmonic model.} Bottom: oscillator modulation depth $k_p$ versus input modulation depth $k_a$. The approximately linear relation \tw{($k_p=0.034~k_a$)} with slope smaller than one confirms the predicted amplitude envelopes attenuation and is in agreement with theory.}
    \label{fig:amp_fit}
\end{figure}

The amplitude response verifies the predicted attenuation of slow envelopes. Figure~\ref{fig:amp_fit} compares normalized envelopes of drive and whistle. The measured output modulation depth $k_p$ scales linearly with the input depth $k_a$, with slope $k_p/k_a \approx 0.034$. Independent measurements of the unforced and forced limit-cycle amplitudes $R$ and $R_F$ give $(1-R/R_F) = 0.028\pm0.016$. This is consistent within uncertainty with the theoretical suppression factor $k_p=(1-R/R_F)\,k_a$ derived in \eqref{eq:k_in_out}.

Taken together, these measurements support the nonautonomous synchronization picture for amplitude modulation. Specifically, the modulation acts as a vertical excursion in the Arnold–tongue plane, enlarging the broad synchronization region through a widened intermittent band while shrinking the persistent core. This behavior is complementary to the broadening but simultaneously to the reduce of the persistent synchronization under frequency modulation and is in quantitative agreement with the slow-flow analysis and its nonautonomous Adler reduction.

\begin{figure}[!htb]
    \centering
    \includegraphics[width=1\linewidth]{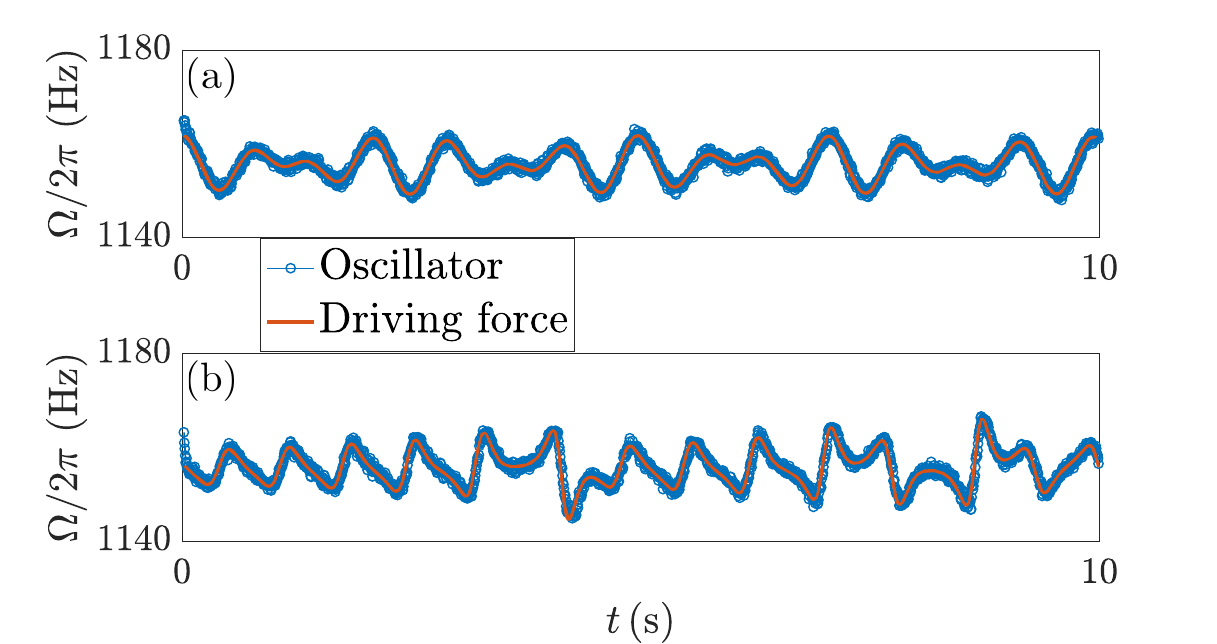}
    \caption{Instantaneous frequency tracking under (a) aperiodic modulation and (b) chaotic‐frequency forcing. For aperiodic forcing, the driving frequency (orange) is modulated by two slow frequencies $\Omega_m$ and $\sqrt{2}\Omega_m$. In the chaotic case, the driving frequency is modulated by a segment of a Lorenz attractor time series, while the oscillator frequency (blue) is plotted in the strict‐synchronization regime. The persistent overlap of the two traces demonstrates robust phase locking despite the arbitrary, chaotic variations in the drive.}
    \label{fig:chaotic}
\end{figure}
\subsection{Aperiodic Modulation}
To probe the limits of phase locking under time variation, the drive frequency is subjected to non-periodic variations and the ability of the whistle to maintain frequency entrainment is examined. In Fig.~\ref{fig:chaotic}(a), the driving frequency (orange) is modulated by two incommensurate sinusoids,
\[\Omega_F(t) = \Omega_u + k_{f,1}\,\Omega_m\sin(\theta_m) + k_{f,2}\,\Omega_m\sin(\sqrt{2}\theta_m),\]
while in Fig.~\ref{fig:chaotic}(b) it follows a normalized segment of a Lorenz time series $s(t)$,
\[\Omega_F(t) = \Omega_u + k_{f}\,\Omega_m\,s(t).\]
In both cases the instantaneous response frequency $\omega_p(t)$ (blue) follows the aperiodically modulated drive $\Omega_F(t)$ (orange) throughout the experiment.

This behavior is consistent with the nonautonomous Adler equation (see \eqref{eq:FM_adler_main}): persistent synchronization for arbitrary detuning trajectories holds whenever the instantaneous locking condition
\[ |\Delta\omega_{\mathrm{eff}}(t)| \le \gamma_{\mathrm{eff}}(t) \]
is satisfied for all $t$. In the present aperiodic tests, parameters are chosen such that \(\sup_t |\Omega_F(t)-\omega_n| < \gamma_0,\) so that the trajectory remains well inside the persistent synchronization region of the Arnold tongue and does not approach the boundaries where saddle-node bifurcations and phase slips would occur.

Figure~\ref{fig:chaotic} shows that, despite quasiperiodic or chaotic detuning, the response frequency overlaps the drive almost perfectly and no phase slips occur over the entire record. These observations are therefore consistent, in the persistent regime, with the theoretical predictions for nonautonomous synchronization in frequency-modulated oscillators.

\section{Conclusion}\label{sec:conclusion}
We present a unified theory–experiment study of synchronization in a self-sustained acoustic oscillator under slowly varying forcing. Previous studies have shown that frequency modulation introduces intermittent regimes and enlarges the broad sense synchronized region, and we find that amplitude modulation produces an analogous widening via a growing intermittent band. Experimental Arnold tongue maps verify the predicted geometries and regime-specific phase slip patterns. The attenuation of slow amplitude envelopes in the amplitude-modulated case is also verified. We further demonstrate persistent tracking under aperiodic (quasi-periodic and chaotic) detuning as long as the instantaneous mismatch remains below the effective coupling. Together, these results validate slow modulation effects for both amplitude and frequency modulation types in an aeroacoustic oscillator. These findings offer practical guidance: frequency modulation enlarges the usable synchronization domain without increasing nominal forcing level, in nonlinear acoustic resonators, tunable metamaterials, synchronized oscillator networks and synchronization-based communications.

\appendix
\section{Frequency–modulated phase: transformation to a nonautonomous Adler equation with time varying detuning}\label{sec:appendix}

Starting from the slow phase dynamics~\eqref{eq:slow_phase_FM},
\begin{equation}
\frac{d\theta}{dT_1}
  =
  \Delta\omega
  - \frac{F\,J_0(k_f)}{2\,\omega_n\,r}\,\cos\theta
  - \frac{F\,J_1(k_f)}{\omega_n\,r}\,
    \cos\bigl(\theta - \theta_m(T_1)\bigr),
\label{eq:app_slow_phase_FM}
\end{equation}
and using the definitions 
\begin{equation}
  \gamma_1 = \frac{F\,J_0(k_f)}{2\,\omega_n\,r},
  \qquad
  \gamma_2 = \frac{F\,J_1(k_f)}{\omega_n\,r}.
\end{equation}
we can write
\begin{align}
\frac{d\theta}{dT_1}
&= \Delta\omega
   - \Bigl[(\gamma_1+\gamma_2\cos\theta_m)\cos\theta
           + \gamma_2\sin\theta_m\,\sin\theta\Bigr] \notag\\
&= \Delta\omega - R(\theta_m)\,\cos\bigl(\theta - \delta(\theta_m)\bigr),
\label{eq:app_Rdelta}
\end{align}
with
\begin{align}
R(\theta_m)
  &= \sqrt{(\gamma_1+\gamma_2\cos\theta_m)^2
            + (\gamma_2\sin\theta_m)^2},\\
\delta(\theta_m)
  &= \arctan \left(
        \frac{\gamma_2\sin\theta_m}
             {\gamma_1+\gamma_2\cos\theta_m}\right).
\end{align}
Introducing the new phase
\begin{equation}
  \psi = \theta - \delta(\theta_m) + \frac{\pi}{2},
\end{equation}
it follows
\begin{equation*}
    \frac{d\psi}{dT_1}
  = \frac{d\theta}{dT_1}
    - \frac{d\theta_m}{dT_1}\,\delta'(\theta_m),
  \quad
  \cos\left(\theta-\delta\right)=\sin\psi,
\end{equation*}
\vspace{0.5ex}

Eq.~\eqref{eq:app_Rdelta} becomes the nonautonomous Adler equation with time-varying detuning,
\begin{equation}
\frac{d\psi}{dT_1}=\Delta\omega_{\mathrm{eff}}(T_1)
  + \gamma_{\mathrm{eff}}(T_1)\,\sin\psi,
\label{eq:app_adler_nonauto}
\end{equation}
where
\begin{align}
\gamma_{\mathrm{eff}}(T_1)
  &= R\bigl(\theta_m(T_1)\bigr),\\
\Delta\omega_{\mathrm{eff}}(T_1)
  &= \Delta\omega
      - \Omega_m\,\delta'\bigl(\theta_m(T_1)\bigr),\\
\delta'(\theta_m)
  &=
    \frac{\gamma_1\gamma_2\cos\theta_m + \gamma_2^2}
         {\gamma_1^2 + 2\gamma_1\gamma_2\cos\theta_m + \gamma_2^2}.
\end{align}
For \(k_f\ll 1\) one has \(J_0(k_f)\approx 1\) and \(J_1(k_f)\approx k_f/2\), hence \(\gamma_2/\gamma_1\approx k_f\) and
\begin{align}
\gamma_{\mathrm{eff}}(T_1)
  &= \gamma_1 + \mathcal{O}(k_f),\\
\Delta\omega_{\mathrm{eff}}(T_1)
  &= \Delta\omega
     - \Omega_m\,k_f\cos\theta_m(T_1)
     + \mathcal{O}(k_f^2).
\end{align}
Therefore, the phase dynamics becomes nearly constant coupling with a periodically modulated detuning, giving the Adler-type phase equation used in the main text.

%%% References
%

\end{document}